\documentclass[jkps,twocolumn,fleqn,showpacs,showkeys]{revtex4}
\usepackage{amssymb}
\usepackage{amsmath}
\usepackage{bm}
\usepackage[dvips]{graphicx,color}

\begin{document}
\setcounter{page}{1}

\title[]{Singlet-triplet Crossover in the Two-dimensional Dimer Spin System YbAl$_3$C$_3$}

\author{Shunichiro \surname{Kittaka}}
\email{kittaka@issp.u-tokyo.ac.jp}
\author{Tomoyoshi \surname{Sugiyama}}
\author{Yasuyuki \surname{Shimura}}
\author{Toshiro \surname{Sakakibara}}
\affiliation{Institute for Solid State Physics, University of Tokyo, Kashiwa 277-8581, Japan}
\author{Saori \surname{Matsuda}}
\author{Akira \surname{Ochiai}}
\affiliation{Department of Physics, Tohoku University, Sendai 980-8578, Japan}

\date[]{Received 4 September 2012}

\begin{abstract}
Low-temperature magnetization ($M$) measurements down to 0.1~K have been performed in magnetic fields up to 14.5~T for 
a single piece of a tiny single-crystalline sample ($\sim 0.2$~mg weight) of the spin-gap system YbAl$_3$C$_3$.
At the base temperature of 0.1~K, several metamagnetic transitions were clearly observed for $H \parallel c$ in the range 6~T $< \mu_0H <$ 9~T
whereas only two transitions were observed, one at 4.8~T and the other at 6.6~T, for $H \parallel a$. 
At fields above 9~T, the magnetization becomes almost saturated for both $H \parallel a$ and $H \parallel c$. 
The present results indicate that a singlet-triplet crossover occurs in a relatively narrow field range, 
suggesting a rather weak interdimer interaction in spite of the nearly triangular lattice of Yb ions.
\end{abstract}



\maketitle

\section{INTRODUCTION}

Low-dimensional quantum spin systems have attracted much interest because of their novel ground states dominated by strong quantum fluctuations.
Intensive studies have been done in the $d$-electron compounds 
such as the two dimensional $S=1/2$ dimer spin systems SrCu$_2$(BO$_3$)$_2$~\cite{Takigawa2010JPSJ} and (CuCl)LaNb$_2$O$_7$~\cite{Kageyama2005JPSJ}, both of which have a singlet ground state.
By contrast, not many $4f$-electron compounds have been investigated from a standpoint of quantum spin systems.
This is because $4f$-electron compounds generally have a large total angular momentum $J$ that is equal to and above $5/2$, which would hinder intersite quantum fluctuations. 
Moreover, in the case of metallic $4f$-electron compounds, either long-range Ruderman-Kittel-Kasuya-Yoshida (RKKY) interactions induce a magnetic ordering or a screening by the conduction electrons leads to a singlet Kondo state.
These features of $4f$-electron compounds tend to disturb the realization of a quantum spin state.
Until recently, only Yb$_4$As$_3$ has been known as a unique $4f$-electron compound in which a one-dimensional pseudo-spin-$1/2$ Heisenberg antiferromagnetic ground state is realized~\cite{Kohgi1997PRB}.

Recently, YbAl$_3$C$_3$ has been proposed to be another candidate for a $4f$-electron quantum spin system. 
YbAl$_3$C$_3$ crystallizes into a hexagonal ScAl$_3$C$_3$-type structure at room temperature, 
which consists of layers of a Yb triangular lattice separated by Al and C layers. 
At temperatures below about 80~K ($=T_{\rm s}$), it exhibits a structural phase transition into an orthorhombic structure, accompanied by a slight displacement of the Yb atoms~\cite{Matsumura2008JPSJ}.
YbAl$_3$C$_3$ was revealed to have a low carrier concentration of about 0.01 per formula unit~\cite{Ochiai2007JPSJ}  
and not to show any long-range magnetic ordering at temperatures down to 0.5~K~\cite{Kato2008JPSJ}.

The magnetic susceptibility $\chi(T)$ of YbAl$_3$C$_3$ in the high-temperature hexagonal phase at temperatures above 80~K 
indicates an effective Yb moment of $4.65 \sim 4.66 \mu_{\rm B}$ and a Weiss temperature of $\Theta=-80\sim-120$~K, 
depending on the field direction, implying the existence of a relatively large antiferromagnetic interaction between the localized Yb$^{3+}$ moments~\cite{Ochiai2007JPSJ}.
In the orthorhombic phase, $\chi(T)$
shows a broad peak around 10~K and becomes quite small at lower temperatures,
suggesting a non-magnetic ground state with the development of a spin gap~\cite{Ochiai2007JPSJ}.
Interestingly, in the specific heat $C(T)$ measurements, a Schottky-type anomaly was found around 5~K, whose entropy release was estimated to be exactly $R\ln2$~\cite{Kato2008JPSJ}.
In addition, inelastic neutron scattering spectra exhibit three well-defined peaks around 1.5~meV, indicating the presence of low-energy magnetic excitations, 
and confirmed that the ground-state Kramers doublet of Yb$^{3+}$ was well separated from the excited levels by about 200~K~\cite{Kato2008JPSJ}.
From these facts, it has been proposed that the ground state Kramers doublet of Yb$^{3+}$, which can be represented by a pseudo-spin 1/2,  forms an antiferromagnetic dimer state with a singlet-triplet energy gap $\Delta$ of about 15~K~\cite{Ochiai2007JPSJ}. 
Indeed, the low-energy spectra in the inelastic neutron scattering experiment can be interpreted by using singlet-triplet excitations~\cite{Kato2008JPSJ}.

For the clarification of the nature of the ground state of the spin-gap system, 
low-temperature magnetization $M$ measurements provide a powerful tool because $M(T,H)$ depends on low-lying magnetic states.
Within an isolated dimer model, 
the application of a magnetic field induces a step-like magnetization as $T\rightarrow 0$
reflecting a change in the ground state from a spin singlet to a triplet.
In addition, in the presence of interdimer interactions and geometrical frustration, the degeneracy of the dimer triplet excited states is removed, and various ground states are expected to appear under magnetic fields.
For instance, the two-dimensional $S=1/2$ dimer spin system SrCu$_2$(BO)$_3$ exhibits magnetization plateaus at 1/8, 1/4, and 1/3 of the saturation magnetization, whose origins have been attributed to the formation of superstructures of the triplet state~\cite{Onizuka2000JPSJ}.

In the case of YbAl$_3$C$_3$, the mechanism of the dimer formation is not so obvious because the displacement of the Yb atoms from the original triangular lattice is very small (only $0.1-0.2$\%)~\cite{Matsumura2008JPSJ}.
Accordingly, one might expect the interdimer interaction to be relatively strong (the dimers are less isolated from each other).
The previous $M$ measurements performed at temperatures down to 1.8~K revealed a broad metamagnetic increase in $M(H)$ that could be interpreted as a singlet-triplet crossover~\cite{Ochiai2007JPSJ}.
In addition, quite recently, $M(H)$ at fields up to 8~T was investigated at about 0.5~K, and multiple metamagnetic steps were found for $H \parallel c$~\cite{Hara2012PRB},
though several pieces of single crystals were used in the experiments.
In order to further examine the nature of the singlet-triplet crossover in magnetic fields, 
we measured $M(T,H)$ for \textit{one piece} of a single-crystalline sample of YbAl$_3$C$_3$ at lower temperatures down to 0.1~K in higher magnetic fields up to 14.5~T.

\section{Experimental}
Single crystals of YbAl$_3$C$_3$ were grown by using an encapsulated tungsten crucible~\cite{Ochiai2007JPSJ}.
The typical weight of the obtained crystal was at most several hundred micrograms.
We measured $M(T,H)$ of a tiny single crystal of YbAl$_3$C$_3$ 
by using a high-resolution capacitive Faraday magnetometer with a vertical field gradient of 8 T/m in a dilution refrigerator~\cite{Sakakibara1994JJAP}.
We recently improved the sensitivity of the measurement by a factor of 100 over the previous apparatus, the details of which will be published elsewhere.

In this paper, we present the data obtained for two samples: sample A (0.16~mg) and sample B (0.23~mg).
Because YbAl$_3$C$_3$ is easily decomposed by a reaction with atmospheric oxygen, 
the sample was wrapped in silver foil with silver paste and then 
fixed firmly on the sample stage of the magnetometer by using GE varnish.
Therefore, a slight misalignment of the crystal orientation may have happened.
In all the data presented, the background magnetization of the magnetometer was subtracted.
For the measurements with $H \parallel a$, we cooled the sample slowly across $T_{\rm s}$ in a magnetic field of 10~T applied along one of the three equivalent $a$ axes, 
so that the orthorhombic phase became a single-domain state.

\section{Results and Discussion}

\begin{figure}
\includegraphics[width=3.2in]{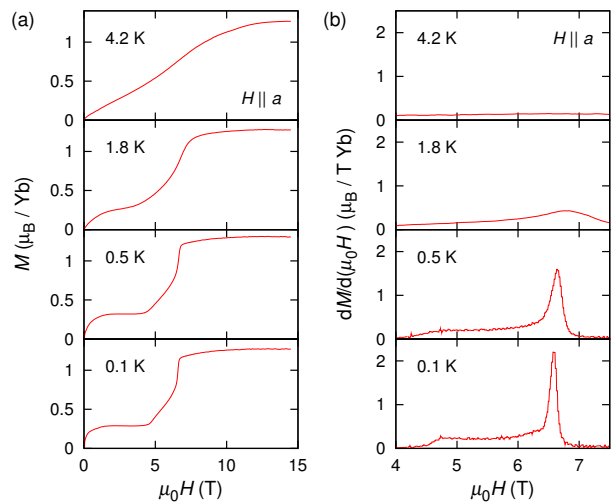}
\caption{(Color online) 
Magnetic field dependences of (a) the magnetization $M$ and (b) the differential susceptibility ${\rm d}M/{\rm d}H$
of sample A for $H \parallel a$ at several temperatures.
}
\label{MH_a}
\end{figure}

Figure~\ref{MH_a}(a) shows the $H$ dependence of the magnetization, $M(H)$, of sample A for $H \parallel a$ obtained at several temperatures.
Whereas $M(H)$ is a gradually increasing function of $H$ at 4.2~K,
the increase in $M$ becomes non-monotonic and steeper at lower temperatures.
At the base temperature of 0.1~K, the differential susceptibility d$M$/d$H$ exhibits a kink at 4.8~T and a large peak at 6.6~T, as shown in Fig.~\ref{MH_a}(b).
Here, no hysteresis was detected down to 0.1~K between the field increasing and decreasing sweeps.
The convex increase in $M(H)$ at low fields below 2~T, which can be fitted using the Brillouin function, 
is attributable to the decomposed Yb$^{3+}$ impurities in the sample.

\begin{figure}
\includegraphics[width=3.2in]{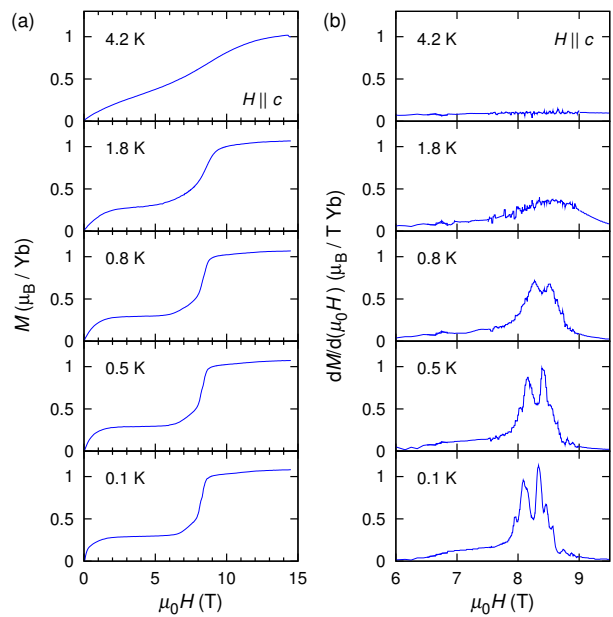}
\caption{(Color online) 
Magnetic field dependences of (a) $M$ and (b) ${\rm d}M/{\rm d}H$
of sample A for $H \parallel c$ at several temperatures.
}
\label{MH_c}
\end{figure}

The application of $H$ along the $c$ axis provides a more striking effect. 
As shown in  Fig.~\ref{MH_c}(a), the gradual increase of $M(H)$ observed at 4.2~K becomes sharp and splits into multiple steps at lower temperatures.
This feature can be seen more clearly in the d$M$/d$H$ data presented in Fig.~\ref{MH_c}(b): 
a kink at around 7~T, which is similar to the one observed for $H \parallel a$ at 4.7~T, and 
more than six peaks in the interval 7.5~T $\le \mu_0H \le 9$~T were observed at 0.1~K.
No appreciable hysteresis was observed for $H \parallel c$, either.
These multiple steps resemble the fractional magnetization steps observed in SrCu$_2$(BO$_3$)$_2$~\cite{Onizuka2000JPSJ},
though clear plateaus cannot be resolved in the $M(H)$ curve for sample~A.

\begin{figure}
\includegraphics[width=3.2in]{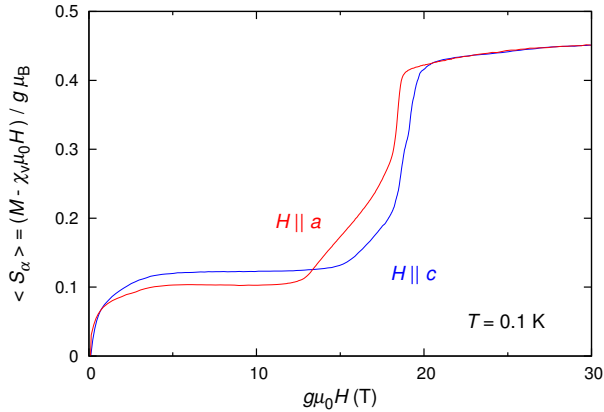}
\caption{(Color online)
Normalized magnetization $\langle S_\alpha \rangle=(M-\chi_{\rm v}\mu_0H)/g\mu_{\rm B}$ of sample A at 0.1~K as a function of $g\mu_0H$.
}
\label{MH_g}
\end{figure}

In a previous report~\cite{Ochiai2007JPSJ},
the $H$ and the $T$ dependences of $M$ of YbAl$_3$C$_3$ were examined at temperatures above 1.8~K and 
were compared by using the isolated dimer spin model with an effective anisotropic $g$ factor. 
In this model, 
the magnetization normalized by the $g$-factor and the Bohr magneton $\mu_{\rm B}$, which is labeled $\langle S_\alpha \rangle$ ($\alpha=a$, $b$, or $c$), is given by~\cite{Kageyama2005JPSJ, Ochiai2007JPSJ}
\begin{align}
\langle S_\alpha \rangle&=\frac{M(H)-\chi_{\rm v}\mu_0H}{g\mu_{\rm B}} \notag \\
&= \frac{N\sinh(g\mu_{\rm B}\mu_0H/k_{\rm B}T)}{1+\exp(\Delta/k_{\rm B}T)+2\cosh(g\mu_{\rm B}\mu_0H/k_{\rm B}T)}. \label{eq1}
\end{align}
Here, $N$, $\chi_{\rm v}$, and $k_{\rm B}$ denote 
the number of Yb ions, the temperature-independent susceptibility, and the Boltzmann constant, respectively.
Accordingly, $\langle S_\alpha \rangle$ is expected to be scaled by $g\mu_0H$;
$\langle S_\alpha \rangle$ increases rapidly and saturates to a value 0.5 at around $g\mu_0H \sim 22.5$~T when $\Delta/k_{\rm B}=15$~K and $T=0.1$~K.

Figure~\ref{MH_g} presents $\langle S_a \rangle$ and $\langle S_c \rangle$ for YbAl$_3$C$_3$ at 0.1~K as functions of $g\mu_0H$, 
where the values of $g$ determined from a previous magnetic-susceptibility study~\cite{Ochiai2007JPSJ} were used, 
and $\chi_{\rm v}$ was adjusted so that $M(H)$ in 5~T $\le g\mu_0H \le 10$~T was almost constant. 
As expected from Eq.~\eqref{eq1}, 
both $\langle S_a \rangle$ and $\langle S_c \rangle$ saturate to be about 0.45 at around $g\mu_0H \sim 20$~T, 
which means that the magnetic moment of the pseudo-spin $S=1/2$ is almost fully polarized.
However, the structures of the magnetization curves do not match between $H \parallel a$ and $H \parallel c$.
For instance, $\langle S_a \rangle$ ($\langle S_c \rangle$) keeps increasing with a gradual slope above $g\mu_0H \sim 13$~T
until it reaches about half (quarter) the saturation magnetization. 
Thus, the $M(H)$ behavior at low temperatures cannot be explained by using the simple isolated dimer model.

The ratio of the saturation fields, $H^\ast_{\rm c2}$ ($=g\mu_0H \sim 18-20$~T), to the onset field of the triplet crossover, $H^\ast_{\rm c1}$ ($=g\mu_0H \sim 13-15$~T),
is estimated to be at most 1.5 for both $H \parallel a$ and $H \parallel c$ in YbAl$_3$C$_3$. 
This ratio is related to the strength of the interdimer interaction.
In the isolated dimer model at 0~K, $H^\ast_{\rm c2}/H^\ast_{\rm c1}=1$.
By contrast, when the interdimer coupling is sufficiently strong, $H^\ast_{\rm c2}/H^\ast_{\rm c1}$ becomes larger 
because the interdimer interaction lifts the degeneracy of the triplet states and makes a wide triplet crossover or induces an ordered state. 
The observed ratio $H^\ast_{\rm c2}/H^\ast_{\rm c1} \sim 1.5$ in YbAl$_3$C$_3$ indicates that the interdimer interaction is not strong compared with other dimer systems, e.g.
SrCu$_2$(BO$_3$)$_2$ ($H^\ast_{\rm c2}/H^\ast_{\rm c1} \gg 3$)~\cite{Onizuka2000JPSJ}, (CuCl)LaNb$_2$O$_7$ ($\sim 3$)~\cite{Kageyama2005JPSJ},
BaCuSi$_2$O$_6$ ($\sim 2$)~\cite{Jaime2004PRL}, and Ba$_3$Cr$_2$O$_8$ ($\sim 2$)~\cite{Kofu2009PRL}.

\begin{figure}
\includegraphics[width=3.2in]{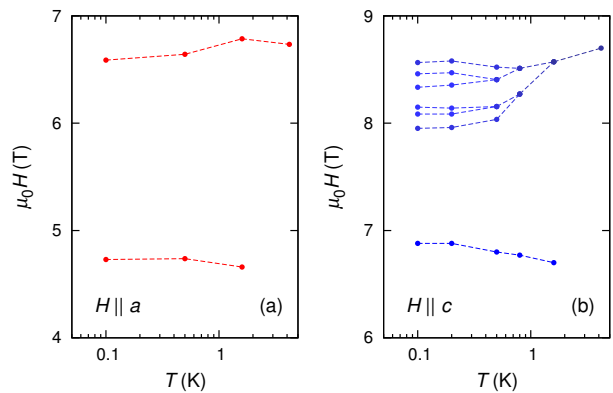}
\caption{(Color online) 
Field-temperature phase diagrams of sample A 
for (a)~$H \parallel a$ and (b)~$H \parallel c$.
Dashed lines are guides to the eye.
The horizontal axis is a logarithmic scale.
}
\label{HT}
\end{figure}

In Fig.~\ref{HT},
we plot the position of a peak or a kink in  d$M$/d$H$ as a function of temperature.
We  define three main regions: a dimer state at low fields, a transition region $H^\ast_{\rm c1}<g\mu_0H<H^\ast_{\rm c2}$, and 
a fully-polarized state above $H^\ast_{\rm c2}$.
Possibly, the transition region consists of a mixture of the singlet and the triplet states.
A remarkable feature is the existence of various internal states in the transition region for $H \parallel c$ at low temperatures represented by the several peaks in d$M$/d$H$.
Because the boundary of the transition region is not likely to close in the $H-T$ plane, we consider that there is no long-range ordered state.

\begin{figure}
\includegraphics[width=3.2in]{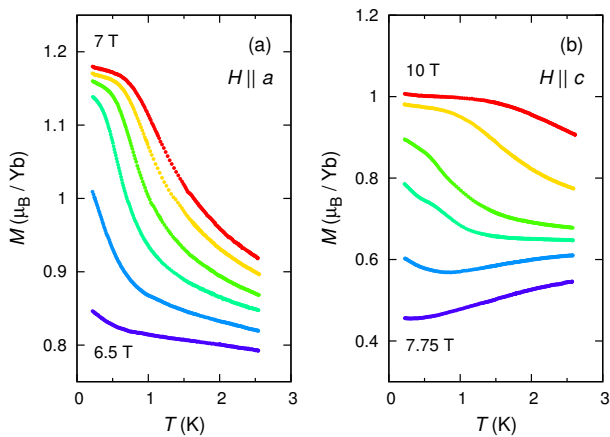}
\caption{(Color online)
Temperature dependences of $M$ of sample A measured (a) in 7, 6.9, 6.8, 6.7, 6.6, and 6.5 T for $H \parallel a$ and
(b) in fields of 10, 9, 8.5, 8.35, 8.1, and 7.75 T for $H \parallel c$, respectively, from top to bottom.
}
\label{MT}
\end{figure}

Figures~\ref{MT}(a) and \ref{MT}(b) show the $T$ dependences of $M$ at several fields for $H \parallel a$ and $H \parallel c$, respectively.
In the field region where $M(H)$ shows a steep increase, a rather strong increase of $M(T)$ was observed on cooling, although there is no distinct feature that could be ascribed to the manifestation of a phase transition.
The stepwise increase in $M(T)$ at 8.35~T for $H \parallel c$ probably comes from the multiple magnetization steps in $M(H)$ for this field direction.

\begin{figure}
\includegraphics[width=3.2in]{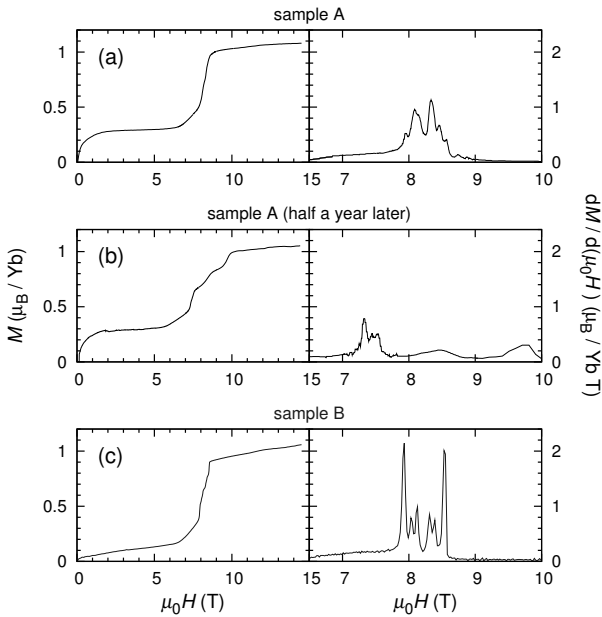}
\caption{
(a)~$M(H)$ and d$M$/d$H$ of sample A replotted from Fig.~\ref{MH_c}, and
(b)~those measured six months later.
(c)~$M(H)$ and d$M$/d$H$ of sample B measured relatively soon after the crystal growth.
All the data presented here have been taken at 0.1~K in fields $H\parallel c$ .
}
\label{sample-dep}
\end{figure}

Next, we focus on the sample quality dependence of the multiple magnetization steps.
After taking the data shown in Fig.~\ref{MH_c}, we kept sample A in a vacuum desiccator.
Six months later, we measured $M(H)$ for sample A again, and the result is shown in Fig.~\ref{sample-dep}(b). 
The data in Fig.~\ref{MH_c} are also replotted in Fig.~\ref{sample-dep}(a) for comparison.
A qualitative difference between the two data can be seen.
In Fig.~\ref{sample-dep}(b), the field range of the transition region becomes wider, and even clear plateaus appear.
The impurity magnetization in low fields is also slightly enhanced, indicating that the sample is degraded and that the number of decomposed Yb atoms has increased.
Obviously, the magnetization process of YbAl$_3$C$_3$ very much depends on the sample quality.
This effect might originate from the partial release of the frustration that induced the distribution of the interdimer interaction.

This, however, does not mean that the observed multiple magnetization steps are caused by such sample deterioration.
Figure~\ref{sample-dep}(c) shows the $M(H)$ for sample B at 0.1~K,
which was measured relatively soon after the growth.
Sample B can be seen to be of good quality because of less impurity magnetization in low fields.
Nevertheless, it shows clear multiple steps with sharp and large peaks of d$M$/d$H$.
In addition, the field range of the transition region for sample B is even slightly wider than that for sample A (Fig.~\ref{sample-dep}(a)).
Note that the largest two peaks in d$M$/d$H$ are well separated for sample B whereas they tend to merge for sample A.
These results strongly suggest that the multiple magnetization steps are intrinsic to the system and are not due to the sample deterioration.
Unfortunately, the $M(H)$ for sample B at different $T$ was not investigated before it had decomposed.

The overall behavior of $M(H)$ in YbAl$_3$C$_3$ is well interpreted by using a singlet-triplet crossover in a spin-dimer system.
In particular, the multiple magnetization steps observed for $H\parallel c$ are reminiscent of the magnetization plateaus seen in the quantum dimer spin system SrCu$_2$(BO$_3$)$_2$, whose origin is the formation of magnetic superstructures of localized triplets.  
It is highly interesting how the dimer state is realized in the nearly triangular lattice of Yb ions in YbAl$_3$C$_3$.
If the singlet-triplet crossover of YbAl$_3$C$_3$ is to be studied in more detail, careful investigations using a high-quality single crystal are essential. 

\section{CONCLUSIONS}
We have measured the magnetization of a single crystal of YbAl$_3$C$_3$ at low temperatures down to 0.1~K and in magnetic fields up to 14.5~T.
For both $H \parallel a$ and $H \parallel c$, a steep increase of the magnetization ascribed to a single-triplet crossover, was observed at 1.8~K.
With decreasing temperature, this crossover becomes sharp and splits into multiple steps.
At the lowest temperature of 0.1~K, multiple magnetization steps,
which resemble the quantized magnetization plateaus found in the quantum dimer spin system SrCu$_2$(BO$_3$)$_2$, were observed only for $H \parallel c$. 
We found that these multiple magnetization steps strongly depended on the sample quality. 
The width of the singlet-triplet crossover region for a good-quality sample was relatively narrow compared with those for other $3d$ electron dimer systems.
This might indicate that the effective interdimer interaction in YbAl$_3$C$_3$ is not so strong in spite of the nearly triangular lattice configuration of Yb atoms.
Further investigations are needed to establish the nature of the novel ground state of YbAl$_3$C$_3$.

\section*{ACKNOWLEDGEMENTS}
This work has been partially supported by a Grant-in Aid for Scientific Research on Innovative Areas (20102007, 23102705)
from the Ministry of Education, Culture, Sports, Science and Technology of Japan, and by a Grant-in-Aid for Scientific Research (A)
 (No. 20244053) from the Japan Society for the Promotion of Science.

\end{document}